\documentclass[a4paper, 12pt, technote, onecolumn]{IEEEtran}

\usepackage[latin1]{inputenc}
\usepackage[T1]{fontenc}
\usepackage[english]{babel}

\usepackage[caption=false,font=footnotesize]{subfig}
\usepackage[pdftex]{graphicx}
\DeclareGraphicsExtensions{.pdf,.jpeg,.png}

\usepackage[cmex10]{amsmath}
\usepackage{trfsigns}

\usepackage{cite}
\usepackage[nolist]{acronym}

\usepackage[detect-all, per-mode=fraction]{siunitx}
\usepackage{booktabs}
\usepackage{multirow}
\usepackage{threeparttable}

\usepackage[version=3]{mhchem}  

\newcommand{\spacepipe}{\,|\,}  
\newcommand{\eqdot}{\qquad .}   
\newcommand{\eqcomma}{\qquad ,} 

\title{A New Frequency Control Reserve Framework\\ based on Energy-Constrained Units}

\author{
  \IEEEauthorblockN{Theodor Borsche, Andreas Ulbig and G\"oran Andersson}
  \IEEEauthorblockA{Power Systems Laboratory, ETH Z\"urich\\
    borsche\spacepipe ulbig\spacepipe andersson\,@\,eeh.ee.ethz.ch
  }
}

\begin{document}
\maketitle

\begin{acronym}[swissgrid]
\acro{LFC}{Load Frequency Control}
\acro{BESS}{Battery Energy Storage System}
\acro{EV}{Electric Vehicle}
\acro{RTO}{Regional Transmission Organization}
\acro{FT}{Fourier Transform}
%
%
\acro{RES}{Renewable Energy Sources}
\acro{PV}{Photo-Voltaic}
%
%
\acro{DSO}{Distribution System Operator}
\acro{TSO}{Transmission System Operator}
%
%
\acro{LV}{Low-Voltage}
\acro{MV}{Medium-Voltage}
\acro{HV}{High-Voltage}
\acro{HVDC}{High-Voltage Direct Current}
%
\acro{BG}{Balance Group}
\acro{BE}{Balancing Energy}
%
\acro{ACE}{Area Control Error}
\acro{AGC}{Automatic Generation Control}
%
\acro{DR}{Demand Response}
\acro{DSM}{Demand Side Management}
\acro{EWH}{Electric Water Heater}
\acro{HVAC}{Heating, Venting and Air-Conditioning}
\acro{PLC}{Power Line Communication}
%
\acro{SOC}[SoC]{State of Charge}
\acro{SOH}[SoH]{State of Health}
%
\acro{MPC}{Model Predictive Control}
\acro{RHC}{Receding Horizon Control}
\acro{LP}{Linear Program}
\acro{QP}{Quadratic Program}
\acro{MILP}{Mixed-Integer Linear Program}
\acro{sLP}[SLP]{Stochastic Linear Program}
%
\acro{PF}{Power Flow}
\acro{OPF}{Optimal Power Flow}
%
\acro{EKZ}{the utility of the Kanton Zurich}
\acro{CAISO}{California Independent System Operator}
\acro{ERCOT}{Electric Reliability Council of Texas}
\acro{swissgrid}{Swiss Transmission System Operator}
\end{acronym}

\begin{abstract}
Frequency control reserves are an essential ancillary service in any electric power system, guaranteeing that generation and demand of active power are balanced at all times. Traditionally, conventional power plants are used for frequency reserves. There are economical and technical benefits of instead using energy constrained units such as storage systems and demand response, but so far they have not been widely adopted as their energy constraints prevent them from following traditional regulation signals, which sometimes are biased over long time-spans. This paper proposes a 
frequency control framework that splits the control signals according to the frequency spectrum. This guarantees that all control signals are zero-mean over well-defined time-periods, which is a crucial requirement for the usage of energy-constraint units such as batteries.
A case-study presents a possible implementation, and shows how different technologies with widely varying characteristics can all participate in frequency control reserve provision, while guaranteeing that their respective energy constraints are always fulfilled.
\end{abstract}

\section{Introduction}
Frequency Control is one of the most important ancillary services in any power system, as it is directly tied to the active power balance. Failure in preventing large frequency excursions can lead to load shedding, generation tripping and, in the worst case, wide-spread blackouts.
Most power grids use a two- or three-level control strategy. The fastest control is the \emph{primary frequency control}, also known as droop or frequency response. It is a distributed, proportional control capable of arresting frequency changes after a change in production or demand. \emph{Secondary frequency control}, also known as \ac{LFC}, employs a central controller called the \ac{AGC}. The \ac{AGC} reacts slower than primary control but has an integral part in the control loop. It is thus able to bring the system frequency back to nominal frequency, depending on the grid region \SI{50}{\hertz} or \SI{60}{\hertz}, as well as to keep exchanges between areas at the scheduled value. While secondary control could also handle lasting generation-demand mismatches, it is desirable to relieve this service to prepare for future disturbances. Depending on the \ac{TSO} and the market structure, different strategies are used. These strategies include activation of tertiary control, spinning and non-spinning reserves as well as the sourcing from intra-day markets. See \cite{Kundur1994} for a detailed technical description of frequency control reserves, and \cite{Rebours2007} for an overview over terminologies and strategies employed by different system operators.

Some energy constrained units, such as batteries, have outstandingly high ramp rates and low or no ramping costs, making them promising candidates for frequency reserves. However, limits on the energy capacity are a challenge for these unit types when following reference signals that are not zero-mean: \ac{SOC} constraints may be hit, rendering the unit unable to continue provision of the agreed upon ancillary service. Current frameworks for frequency control reserves assume conventional power plants\cite{ENTSO-E2009P1}, and therefore do not explicitly take into account or regulate how energy constraints should be handled. 

As of today, frequency control reserves are adequate and offer a high level of robustness against disturbances and contingencies. Nevertheless, there may be benefits associated with using energy constrained units for frequency control reserves. These benefits are 1) of economic nature, 
as it allows for a more efficient unit commitment in many situations by effectively decoupling energy provision from reserve power provision,
 and 2) of technical dimension, as units with very fast ramping ability can improve frequency stability in grids with low rotational inertia and uncertain in-feed from renewable energy sources. 
Both issues will be discussed in detail in Section~\ref{sec:review}.

Past research was looking at methods to provide frequency reserves using energy constrained units in the current frequency reserve framework and was most concerned with the technological feasibility. However, most solutions propose workarounds or add-ons to allow for operation of energy constrained units within the existing framework. 
Some authors, explicitly or implicitly, employ band-pass and high-pass filters for the \ac{AGC} signal to allow a specific type of storage or load to participate in the control reserve markets. The contribution of this paper is to propose a generalized framework for frequency control reserves, which extends the idea of filtering to all parts of the frequency spectrum and to both primary and secondary control.
The advantages of different storage technologies, be it fast ramp rates, high efficiency or cheap storage capacity, can then be harnessed within this framework in a straight-forward and simple manner.

The paper is organized as follows: Section~\ref{sec:review} gives a rough overview of past research on energy constrained units for frequency control. Section~\ref{sec:statusquo} analyses the status quo, before Section~\ref{sec:framework} proposes adjustments to the current control reserve framework. Finally, Section~\ref{sec:casestudy} shows in a case study how the system might work.

\section{Economical and Technical Benefits}
\label{sec:review}
\subsection{Pilot Projects and Economic Benefits}\label{sec:decouple}
Using batteries for primary frequency control was already proposed in the early 1980's. 
The electric utility company of West-Berlin, BEWAG, had to operate this city of about 2 million people as one large island grid throughout the Cold War era. At the time, it was found to be economically advantageous to use a combination of batteries and a base-load plant, rather than to invest in a more expensive and less efficient plant with load-following capabilities\cite{Kunisch1986}. Similar results were found in northern Chile, where the installation of a \SI{12}{\mega\watt} \ac{BESS} for emergency frequency control allows to operate generators closer to maximum capacity and thus to increase their continuous output by \SI{14}{\mega\watt}, while at the same time increasing the mean frequency response from \SI{6}{\mega\watt} to \SI{12}{\mega\watt}\cite{Hsieh2012}.

While the benefit in Chile is to better utilize existing conventional production capacity, a different effect can be observed in Switzerland. To offer reserves, both conventional and hydro power plants have to produce a certain base amount of power. More specifically, the set point has to be at least equal to the offered reserve capacity plus the minimum load at which the plant can run. At hours with low prices, fuel costs or water costs may be higher than what is earned at the power markets. Accordingly, the remuneration for the reserve capacity has to be sufficiently high to also cover the opportunity cost associated with the technically necessary but economically loss-making energy production. When offering reserves with batteries, such a base production is not necessary -- in other words, control reserve \emph{power} is decoupled from \emph{energy} production.

This decoupling of power and energy is the main economic benefit of using storage systems to offer control reserves. Assume, that market prices are low due to high in-feed from renewable source. If conventional power plants are still used to offer frequency control reserves, overall system \ce{CO2} emission will increase as these plants' energy production would not necessarily be needed.
In essence, the unit commitment will at times be suboptimal due the additional constraint of enforcing must-run generation to stay online. Also, in a situation where more than \SI{100}{\percent} of the load is covered by renewable in-feed, one might tend to simply down-regulate wind or \ac{PV} plants so that they can offer both up- and down control reserves. But this would reduce the mean output and waste a certain percentage of clean energy. Instead, it might be more profitable to store excess energy in a storage with slow dynamics utilizing all of the produced energy, and have a storage with fast dynamics provide frequency reserves.

\subsection{Technical Benefits}\label{sec:technical}
\subsubsection{Technical Implementation}
As mentioned in the introduction, the uncertainty about the energy content of frequency control reserve activation patterns makes it very difficult for units with energy constraints to offer frequency reserves. Considering primary frequency control, the energy content of the signal is so small, that power plants are only remunerated for power capacity and not for the actual energy delivery. Still, as long as no guarantees are given, which would be rather difficult in the current frequency control framework, the frequency may be biased over prolonged periods of time, and units participating in the service have to be able to fully follow the signal. Additionally, long tendering periods pose an additional challenge. On the other hand, a stable power system operation is paramount and it is therefore not feasible to allow less strict rules, and shorter tendering periods, although they would add flexibility to the reserve procurement process, might increase both administrative costs and increase volatility.

Approaches to these challenges are manifold. Some authors conclude that the amount of reserve power that can be reliably offered with \ac{DR} is high for short time periods but small for longer durations and argue for a reduction of tendering periods\cite{Koch2009}. Others use stochastic optimization with probability constraints, and then give guarantees on the ability to follow any future signal. Doing so significantly increases the amount of reserves that can be offered, but also introduces a level of uncertainty. Furthermore, pooling of energy constrained units with a conventional power plant is proposed. Thus large reserves can be offered and at the same time the flexibility of, e.g., electric vehicles can be harnessed\cite{Jin2013, Galus2011}. However, it does not explicitly offer a decoupling of power and energy provision -- which we argued in Section~\ref{sec:decouple} is favourable from a functional and optimization perspective -- and it requires the aggregator to own or contract a power plant. Recent studies investigating \acp{EV}, which have a substantial storage capacity and can also feed in to the grid, concluded that reserves can be offered -- but even here, tendering periods limit the amount of reserves, as there are times when only some vehicles are connected to the grid\cite{Ulbig2010IREP, Gonzalez2013}. Finally, \cite{Hao2013ACC} and \cite{Lin2013} propose using \ac{HVAC} systems of large commercial buildings to offer frequency regulation. To ensure continuous provision of the service and avoid interference with the actual building control while limiting the effect on indoor climate, the regulation signal is filtered with a band-pass.

Recent research in primary frequency control reserve provision by batteries investigated different recharging strategies. The pilot project in West Berlin used recharging during low-load hours, and the authors found that it is sufficient to do so three times a week\cite{Kunisch1986}. Oudalov et. al. propose recharging during periods when system frequency is in the dead-band around nominal frequency, and showed that this is sufficient considering one month of historic data\cite{Oudalov2007}. However, no guarantees on the ability to follow are inherent to this system. The authors in \cite{Borsche2013GM} propose to subtracts a moving average from the control signal, thus ensuring that it is zero-mean and the required energy capacity is limited. At the same time, the battery closely follows the fast dynamics of the regulation signal. In \cite{Megel2013}, a strategy activating recharging when the \ac{SOC} reaches a pre-set limit is investigated, and it is shown that this strategy leads to lower cycling of energy than the one proposed in \cite{Borsche2013GM}. However, due to the non-linear behaviour around the \ac{SOC} limits, the control is somewhat chaotic -- following the same frequency signal might trigger a recharging in one case, but not in another, simply due to small differences in \ac{SOC} estimation.

\ac{DR} has also been proposed for frequency response services. For exmaple, \cite{Xu2011} describes a system where loads with thermal storage are automatically disconnected when the system frequency drops by more than \SI{100}{\milli\hertz}, thus helping to stabilize the grid. This approach offers much promise for actual implementation, but avoids the issues of \ac{SOC} constraints as the regulation signal is only being followed in extreme cases.

Secondary control as a centralized control scheme has seen modifications to allow a wider range of units to provide control reserves. Notably, PJM, a \ac{RTO} responsible for a large part of the transmission system in the eastern part of the US, offers a modified signal. Here, the strategy is to split the signal in a high-frequency part, termed RegD signal, and a slower part, termed RegA\cite{PJM2013}. Remuneration for following the faster RegD signal is significantly higher than for the RegA signal.
A similar approach is currently being investigated by the Swiss \ac{TSO} swissgrid\cite{Avramiotis2014}.

\subsubsection{Benefits for System Operation}
Batteries are widely used in micro-grids and island systems to provide frequency control reserves. Optimal sizing of a \ac{BESS} for an island grid is investigated in \cite{Mercier2009}, and specifically notes the improvement of system frequency, both considering peak frequency and settling time, that is achievable by using fast-responding batteries for frequency control.
While island systems with small load and fluctuating in-feed from \ac{PV} or wind might seem a rather specific case, much can be learned from these also for operation of large interconnected power systems. Renewable energy sources have a two-fold effect on system stability: not only does the stochastic production require control reserves, but the fact that most wind and basically all PV plants are connected via inverters does influence rotational inertia of the power system. The reduced rotational inertia leads to faster and larger frequency deviations after a contingency\cite{Ullah2008}. 
If today's frequency control reserves are not able anymore to counteract these more rapid frequency dynamics, this in turn can lead to frequency deviations exceeding acceptable limits -- a very similar effect to those observed in small island grids. Very fast responding units such as batteries might be a solution for this challenge

\section{Analysis of the Status Quo}\label{sec:statusquo}
In this section, we will analyse the primary and secondary regulation signals in more detail. Specifically, we will have a look at the \ac{FT} of the regulation signals. It is very important to distinguish between the system frequency, which refers to the actual rotational speed of generators in the system, and the frequency spectrum of this signal. If we refer to low or high \emph{frequencies}, this refers to the frequency spectrum, while low or high \emph{system frequency} refers to a situation where system frequency deviates from nominal frequency. Also, the \emph{amplitude frequency response} denotes the behaviour of a certain control loop when faced with a signal of a certain frequency, rather than the activation of, e.g., the primary frequency response control service.

\subsection{Amplitude Frequency Response of Regulation Signals}
\begin{figure}
  \centering
  \includegraphics[scale=1]{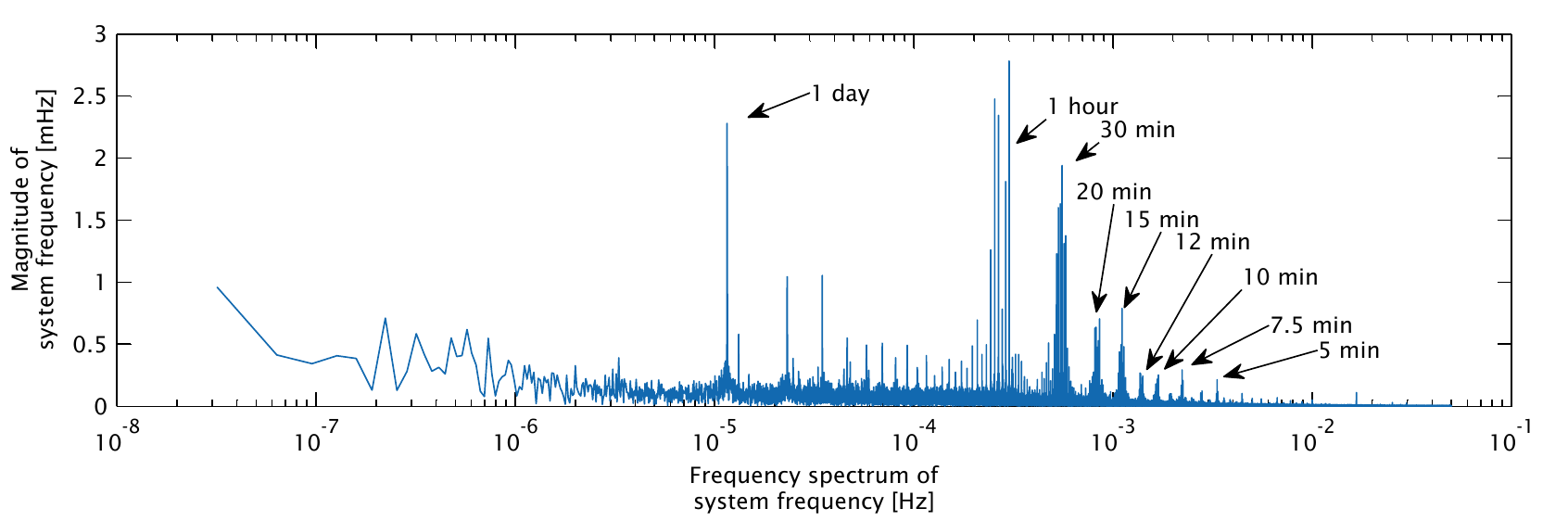}
  \caption{Spectrum of the System Frequency}
  \label{fig:freq}
\end{figure}
\begin{figure}
  \centering
  \includegraphics[scale=1]{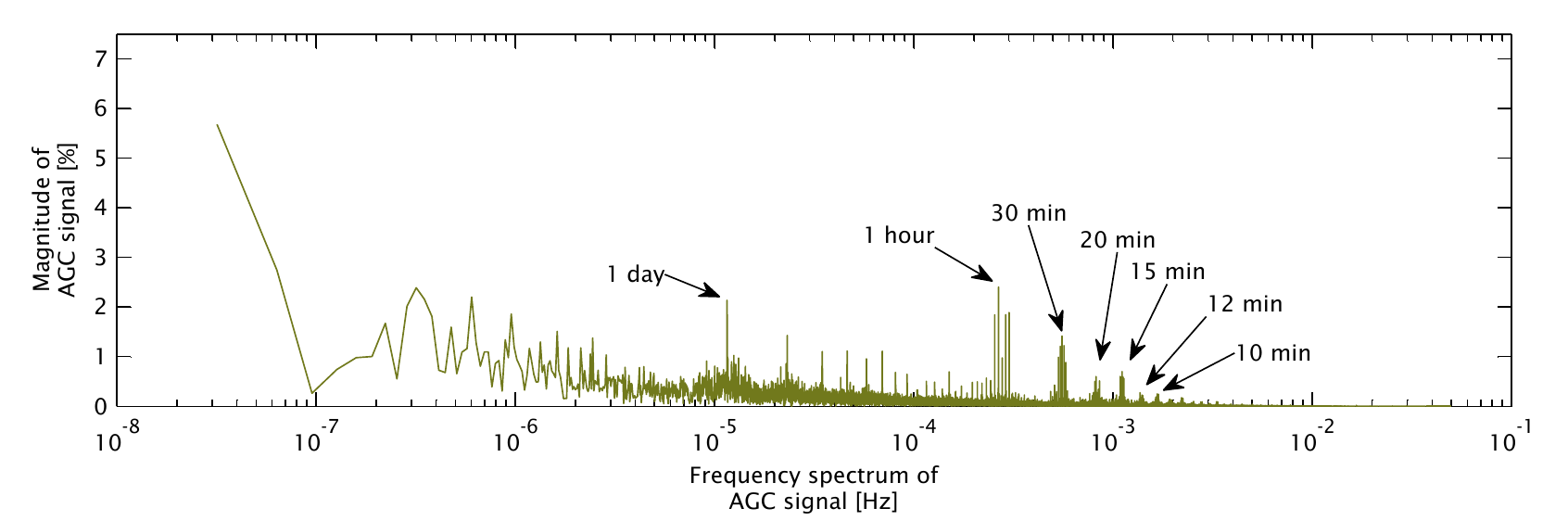}
  \caption{Spectrum of the AGC signal}
  \label{fig:AGC}
\end{figure}
Figure~\ref{fig:freq} shows a Fourier analysis of the system frequency of the continental European grid. Data is in \SI{10}{\second} resolution, covering a period of one year. Certain characteristic effects are of interest. First of all, higher frequencies are well dampened. This is an effect of the inertia of generators. Second, there are several distinct peaks. Annotations in the figure designate the period of these modes. These peaks are artefacts created by the market structure. While the load usually follows a sloped curve, markets are run in hourly and quarter-hourly blocks.  This leads to a repeating sawtooth-shaped mismatch, see also \cite{Weissbach2009PSCE}.
Remember, that the spectrum of a sawtooth wave not only contains a peak with the frequency of the wave itself, but also smaller peaks at multiples of this frequency. We assume, that the peaks with a period of sixty, thirty, twenty, twelve and ten minutes are induced by the hourly blocks in day-ahead markets, while peaks at fifteen, seven-and-a-half and five minutes result from the intra-day markets. Low frequencies are also not very prominent. Keep in mind that the figure represent a system with active control services, meaning that secondary control handles longer lasting deviations.

A similar behaviour can be observed for \ac{AGC} activation, see Figure~\ref{fig:AGC}. While peaks in the middle part of the spectrum are congruent with system frequency, higher frequency peaks are basically non-existent -- which results from the low-pass filter effect of the \ac{AGC}. On the other hand, low frequency peaks are much more pronounced, as secondary control is responsible to bring back system frequency to nominal values and provide power as long as the mismatch persists and tertiary control is not activated.

\subsection{Amplitude Frequency Response of Control Reserves}
Analogue to the analysis of the control signals, the amplitude frequency response of the different control services can be investigated.

The inertia couples frequency change and power mismatch in the system according to the swing equation, here given in the frequency domain and neglecting load damping
\begin{equation}
  \Delta f = \frac{1}{s} \frac{f_0}{2 H S_\mathrm{B}} \Delta P \quad\Leftrightarrow\quad  \Delta P = s \frac{2 H S_\mathrm{B}}{f_0} \Delta f \eqdot
\end{equation}
While this equation usually is interpreted as system frequency increasing or decreasing due to a change in production or load, it also describes the effect that power output of a generator is increased or decreased proportionally to the
rate of change of system frequency $ \Delta \dot{f}(t) \,\laplace\,s \Delta f$ . The generator inertia is therefore the first level of frequency control reserve, and it has the characteristics of a differential controller.

Primary frequency control is a proportional control with the system frequency deviation $\Delta f$ as input. There is no dependency on the frequency rate of change $\Delta \dot{f}$, i.e., the control response is equal regardless of how fast or slow the system frequency changes
\begin{equation}
  P^\mathrm{prim} = -\frac{1}{S} \Delta f \eqdot
\end{equation}
The magnitude of the response is therefore flat. This is true as long as ramp rate constraints and response times of power plants dynamics are neglected: due to these limitations, primary control effectively has a low-pass behaviour with a cut-off frequency around \SI{1e-2}{\hertz}.

Secondary frequency control is a centralized control scheme. Usually, a PI-controller is implemented as \ac{AGC}. For a one area system it may take the form
\begin{equation}
  P^\textrm{sek} = - B \left(  C_\mathrm{p} + \frac{1}{T_\mathrm{N}s} \right) \Delta f \eqdot
\end{equation}
The proportional part has the same effect as the primary frequency control, but with significantly smaller amplitude as $C_\textrm{p} \ll 1$. The integral part is designed to bring the system frequency back to nominal values by cancelling out lasting generation-demand mismatches. In a bode plot, this is seen by an increasingly strong response to low frequencies. $B$ usually is chosen as $\frac{1}{S}$, which in multi-area systems has the advantageous properties of non-interactive control.

Finally, tertiary control is not automatically activated --  an analogous analysis is therefore only qualitatively possible. It can be assumed that tertiary control is used to relieve secondary control, which would be equivalent to a low pass filter with steep flanks -- but generally activation depends on many more system parameters.

\begin{figure}
  \centering
  \includegraphics[scale=1]{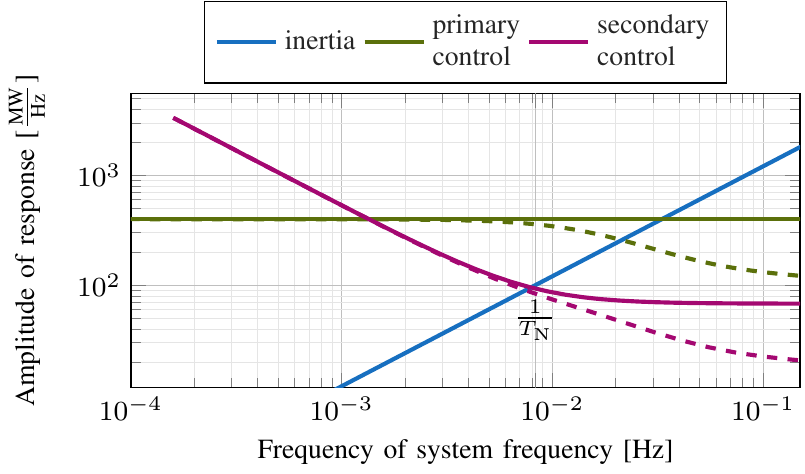}
  \caption{Bode diagram of current control framework. Different services are the main respondents to different parts of the spectrum -- inertia damps high frequency disturbances, while secondary control counteracts low frequency and lasting mismatches. Dashed lines show response when thermal power plant dynamics are taken into account.}
  \label{fig:bode}
\end{figure}

Figure~\ref{fig:bode} shows the different services in one bode plot. For this qualitative plot, the droop is set to \SI{400}{\mega\watt\per\hertz}, AGC parameters $T_\mathrm{N}$ and $C_\mathrm{p}$ are chosen as \SI{120}{\second} and 0.17, respectively, and inertia is assumed to be \SI{6}{\second} with the base-load at \SI{8}{\giga\watt}. These values are typical for the Swiss system. Dashed lines show power plant activation with some dynamics taken into account, namely a reheater with time-constant $T_\mathrm{RH}$ of \SI{10}{\second} and a delay, modeled as low-pass filter, between valve activation and high-pressure turbine $T_\mathrm{CH}$ of \SI{0.3}{\second}. Compare this also to the amplitude plots of system frequency, Figure~\ref{fig:freq} and AGC, Figure~\ref{fig:AGC}.

For frequencies below \SI{1.33e-3}{\hertz}, that is a period of 12 minutes or more, secondary control is the dominant control service. Signals with frequency higher than \SI{0.033}{\hertz}, or a period less than \SI{30}{\second}, are mainly dampened by inertia. This time happens to coincide with the activation time allowed for primary frequency response. In between \SI{1.33e-3}{\hertz} and \SI{0.033}{\hertz}, primary frequency control is the dominant control reserve.

\section{Frequency bands of Frequency Control Reserves}\label{sec:framework}
All recharge strategies described in Section~\ref{sec:technical} can be seen as high-pass filters of some sort, ensuring that the regulation signal is zero-mean over a certain time, or in a less strict interpretation that the integral of the signal and thus the energy content is bounded and finite. 
We also found, that the system frequency deviations have a characteristic frequency spectrum. Finally, the currently implemented control reserve framework has a very characteristic amplitude frequency response.
Considering all this: why not define services that are a-priori limited to certain bands of the frequency spectrum? As mentioned before, such a system is implemented in the PJM interconnection specifically for the secondary control signal\cite{PJM2013}. We aim to derive a framework, where this approach is applied to all bands of the spectrum.

There are some properties of the current control structure that need to be kept in mind, as they are essential for the robustness of frequency reserves.
The integrating behaviour of secondary control must be kept in order to ensure that system frequency and tie-line exchange powers are brought back to nominal values. However, such a control can only be implemented using a centralized control scheme -- or at least a control scheme with exchange of information between the participants.

On the other hand, one of the advantages of primary control is its distributed topology. No communication is needed, and the primary frequency response is thus independent of failures of any communication system, making it rather robust in practical terms. If a new primary frequency control is restricted to high frequencies and a new secondary control to low frequencies, communication outage might lead to serious problems: as the AGC is no longer active, and the new primary control would not arrest lasting power mismatches and therefore not prevent a continuing frequency decay, the system would soon be unstable. In case of loss of communication, the units providing slow control reserves following a central signal would need to \emph{fall-back} to provide a service analogous to primary control.

Finally, tertiary control would need to be activated in such a way that the \ac{AGC} signal is guaranteed to be also zero mean over a predefined time interval. While this could be done by simply averaging over the recent signal and adjusting tertiary output every now and then, it could also be implemented using a optimal predictive control policy, with constraints ensuring that the energy in the AGC signal is bounded. Such predictive methods were already successfully applied to an economic activation of tertiary control to preemptively relief secondary control for the well-known phenomenon of deterministic frequency deviations at the hourly change, caused by power market activity\cite{Farzaneh2012}.

In principle, it would also be possible to completely replace the tertiary control by an intra-day market, with the \ac{TSO} placing bids according to reserve needs -- however  the interplay of tertiary control and intra-day markets depends mainly on the liquidity of the intra-day market. This is not part of this research, but could be easily worked into the proposed framework.

Note, that it would also be possible to include the provision of rotational inertia as an ancillary service in this framework. Inertia can be seen as part of the distributed proportional control, for highest frequencies of the spectrum.

\subsection{Control Power and Balancing Energy Pricing Based on Frequency Bands}
Pricing of services could also be adjusted to frequency bands. While a unit providing the whole spectrum should get similar reimbursement as today, there might be different pricing for different parts of the spectrum. Fast responding units would probably be able to receive high payments for offered control power, while slow ramping units could still participate in the market but would only see comparatively low payments, as is done in the PJM ancillary service market today. This could give an incentive to install and market storage technologies with high ramping capabilities, such as Li-Ion batteries, super-caps or flywheels.
Energy payments would be restricted to the slowest service, that is tertiary control or power from intra-day markets, as all faster services are zero-mean, meaning that faster services are all being replenished eventually by the slower services with only the slowest service delivering a real energy provision. Depending on the amount of available dispatchable energy production capacity, energy might yield very high or low prices -- but importantly, energy prices would be decoupled completely from control power reimbursements.

Vice versa, \acp{BG} could be charged according to the spectrum of their deviations. A Fourier analysis of the daily mismatch between schedule and consumption, measured in sufficient time resolution, would indicate what kind of service was required by the specific balance group. E.g., if units able to provide regulation services with a period of one hour are very expensive, balance groups introducing mismatches with the according frequency could be charged more. This might promote actions by \acp{BG} in scheduling or real-time control that reduces specific types of disturbances, rather than the rather crude punishment of power trajectory deviation by using the energy mismatch over a time-interval of 5 or \SI{15}{\minute} as the relevant metric, as is standard today. However, reducing the deviation between the actual and the scheduled power trajectory of a BG would require significant measurement capabilities, especially when network operation and energy provision are unbundled even on distribution level, meaning that each consumer would have to be measured individually. At the same time, such a scheme would only improve social welfare if \acp{BG} actually have the ability to influence their deviations.

\subsection{One Possible Implementation Scheme}
In the following we will develop a specific implementation of a control framework based on the recharge strategy in \cite{Borsche2013GM} guaranteeing that all regulation signals are zero-mean except the slowest service in the cascade.
First, let us recapture the basic algorithm. Let $P^\mathrm{prim}$ be the primary control signal, and $P^\textnormal{bat}$ the actual output of the battery. The signal then has an offset which can be parametrized by $a$ and $d$, 
\begin{align}
    P^\textnormal{off}(k+d) &= -\frac{1}{a}\sum_{j=k-a}^{k} P^\textnormal{1}(j) \eqcomma\label{eq:offset}\\
  P^\textnormal{bat} &= P^\mathrm{prim} + P^\textnormal{off} \label{eq:batt}\eqdot
\end{align}
Losses can also be taken into account explicitly, see \cite{Borsche2013GM}. In the following, we will assume $d$ to be zero, however using $d$  offers additional flexibility if start-up times or delays of slower units have to be taken into account.

\subsubsection{Distributed, Proportional Control Reserves}
Figure~\ref{fig:bode} shows how different control reserves are active in different domains of the frequency spectrum. It is therefore straightforward to argue, that units providing a distributed, proportional control reserve are only needed in a certain part of the spectrum. 

Following a step input, the above formulation with $d = 0$ and $a = \SI{900}{\second}$ would lead to an instantaneous response, which then linearly decays over a period of \SI{900}{\second}. \cite{Borsche2013GM} argued, that this ramping-down is acceptable as slower services can easily follow this ramp in opposing direction and thus take over from a battery. Beside this handing over responsibility from fast to slow services after the initial response, it would also be possible to split the signal into more parts. Lets assume a storage technology with very small energy capacity but fast ramping and high cycle-lifetime, as well as a slightly bigger storage which also ramps fast but would prefer to avoid the very fast oscillations. Technologies in this thought experiment could be super-caps and Li-Ion batteries, respectively. Their respective behaviour $P^\textnormal{sc}$ and $P^\textnormal{bat}$ can be computed according to
\begin{align}
  \begin{split}
  P^\textnormal{sc}(k)  = &P^\mathrm{prim}(k) - \frac{1}{a^\textnormal{sc}} \sum_{j=k-a^\textnormal{sc}} ^{k} P^\textnormal{1}(j) \eqcomma\\
  P^\textnormal{bat}(k) = &P^\mathrm{prim}(k) - P^\textnormal{sc}(k) - \\
                          &\frac{1}{a^\textnormal{bat}}\sum_{j=k-a^\textnormal{bat}}^{k} \left(P^\textnormal{1}(j) - P^\textnormal{sc}(j)\right) \eqcomma
  \label{eq:distributed_split}
  \end{split}
\end{align}
with $a^\textnormal{bat} > a^\textnormal{sc}$. More levels could be easily added according to specific needs and available storage technology. Two things should be noted: in this formulation, the response would be much faster than current primary frequency response, which has a full activation time of \SI{30}{\second}. Second, lower frequencies of the spectrum are not being followed.

\subsubsection{Centralized, Integrating Control Reserves}
Accordingly, the secondary control signal can be split into several parts. Above examples assume the offset to be adjusted at every time-step, leading to a continuous adjustment of the power that is transferred to slower control services. For units that are currently providing tertiary control, this continuous change of set-point might be undesirable. However, the average could also be sampled on a \SI{15}{\minute} or hourly basis.

\section{A Small Case Study}\label{sec:casestudy}

%
%
%
\begin{figure}
  \centering
  \includegraphics[scale=1]{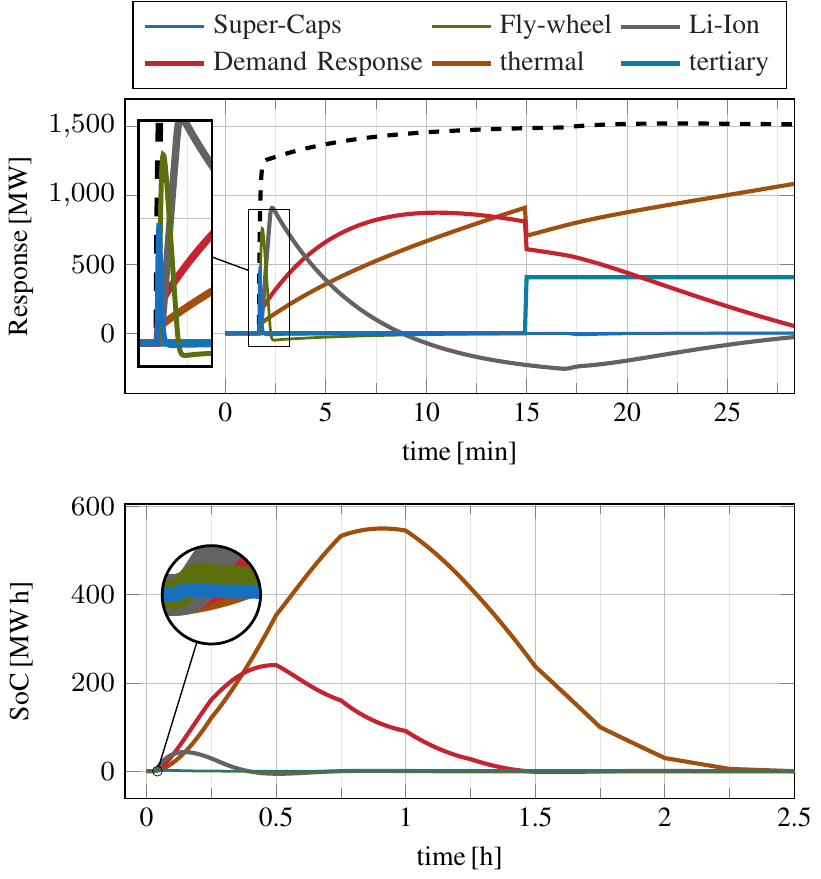}
  \caption{Results of the Case Study. Top: Activation of different control reserves. Fast reserves are activated first and are quickly relieved. Recharging of the battery (grey) leads to a negative power consumption, start of tertiary control or intra-day units (light blue) are scheduled to fixed quarter hours. Total activation is according to system needs (dashed black). Bottom: \ac{SOC} evolution of the different control reserves. After some time, all reserves are brought back to the original SoC. Fast units have minimal energy capacity requirements, the change in SoC cannot be seen.}
  \label{fig:CaseStudy}
\end{figure}

\begin{figure}
  \centering
  \includegraphics[scale=1]{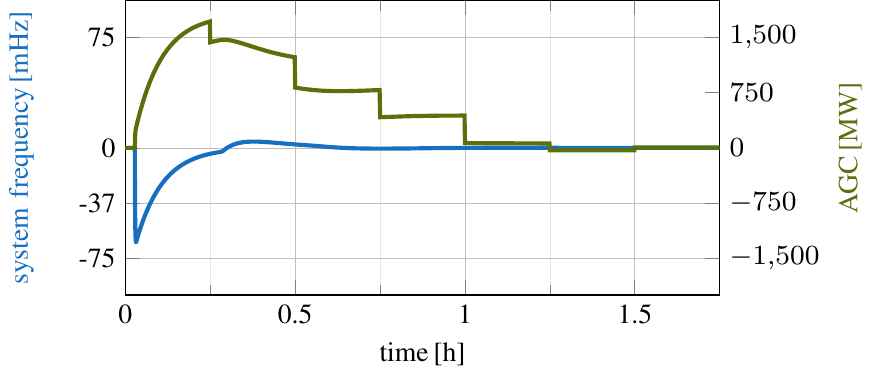}
  \caption{Evolution of system frequency and AGC. Frequency is rapidly brought back to nominal values. AGC activation is relieved by tertiary control in due time.}
  \label{fig:CaseStudy_reg}
\end{figure}

A small case study is defined to illustrate the proposed framework. Frequency control reserves from a pool of different technologies are used. For the distributed proportional control super-caps, flywheels and Li-Ion-batteries are chosen, and for the AGC demand response and a thermal power plant. Distributed control is active only for high frequency parts of the spectrum, being relieved after \SI{900}{\second}, while the central control only handles low frequencies with a maximum ramp rate of \SI{300}{\second} for full activation. Secondary control is relieved with power from intra-day markets that can, as assumption, be sourced at every quarter hour.

The reserves providing fast response further split the frequency spectrum according to their ability. Super-caps are relieved after $a^\textnormal{sc} = \SI{5}{\second}$, fly-wheels after $a^\textnormal{fw} =\SI{30}{\second}$ and batteries after $a^\textnormal{bat} =\SI{900}{\second}$. Some indices are omitted for brevity, the detailed formulation is given in \eqref{eq:distributed_split}. The different units then follow these signals
\begin{align}
  P^\mathrm{sc} &= P^\mathrm{prim} - \frac{1}{a^\textnormal{sc}} \sum P^\mathrm{prim} \\
  P^\mathrm{fw} &= P^\mathrm{prim} -P^\textnormal{sc} - \frac{1}{a^\textnormal{fw}} \sum \left( P^\mathrm{prim} -P^\textnormal{sc} \right) \\
  P^\mathrm{bat} &= P^\mathrm{prim} - P^\textnormal{sc} -P^\textnormal{fw} - \frac{1}{a^\textnormal{bat}} \sum \left( P^\mathrm{prim} - P^\textnormal{sc} -P^\textnormal{fw} \right)
\end{align}
Note, that the slower units do not need to know the power output of the fast units -- it would be sufficient to know the design parameters, which are fixed beforehand, and which deterministically  govern the activation of the other units.

Considering the units providing central integrating control, we assume that they can be activated within five minutes. \ac{DR} can handle changes of consumption for up to \SI{30}{\minute} before starting to recharge, and provides \SI{70}{\percent} of the reserve during this time. The thermal power plant participating is more expensive for fast actions and its bid in an imagined ancillary service auction was successful for \SI{30}{\percent} of the reserve power in this specific frequency band. Lower frequencies are completely handled by the thermal plant. Finally, intra-day markets are used to both relieve the secondary control, and to guarantee that the activation of all units is zero-mean with respect to their respective time spans. Energy reserves are activated by averaging AGC activation over \SI{1}{\hour}. While energy might not be a constraint for the thermal plant, the case study will show that the zero-mean condition holds for all units. The activation follows these rules
\begin{align}
  P^\mathrm{DR} &= 0.7 \left( P^\mathrm{AGC} - \frac{1}{a^\textnormal{DR}} \sum P^\mathrm{AGC} \right) \\
  P^\mathrm{therm} &= P^\mathrm{AGC} - P^\mathrm{DR} - \frac{1}{a^\textnormal{therm}} \sum P^\mathrm{AGC} - P^\mathrm{DR} 
\end{align}

\begin{table}
  \centering
  \caption{Simulation Parameters}
  \label{tab:sim}
  \begin{tabular}{l l r@{\,}l}
    \toprule
    parameter                & variable               & \multicolumn{2}{c}{value} \\ \midrule
    rotational inertia       & $H$                    & 6    & \si{\second}    \\
    base power               & $S_\textnormal{B}$     & 280  & \si{\giga\watt} \\
    Primary Control reserves & $P^\textnormal{prim,max}$ & 3000 & \si{\mega\watt} \\
    droop                    & $1/S$                  & 15000  & \si{\mega\watt\per\hertz} \\
    Secondary Control reserves & $P^\textrm{sek,max}$ & 15000  & \si{\mega\watt}  \\
    AGC parameters           & $C_\textnormal{p}$     & 0.17 \\
                             & $T_\textnormal{N}$     & 200  & \si{\second} \\
    Load-frequency damping   & $D_\textnormal{l}$     & $\frac{1}{4200}$& \si{\hertz\per\mega\watt}  \\ \bottomrule
  \end{tabular}
\end{table}

\begin{table}
  \centering
  \caption{Simulation Results}
  \label{tab:results}
  \begin{tabular}{lS[table-format=-3.2]S[table-format=-4.2]S[table-format=-3.2]S[table-format=-1.2]S[table-format=-3.2]}
    \toprule
     & \multicolumn{2}{c}{$P$} & \multicolumn{2}{c}{SoC} & {$E^\textnormal{cycled}$} \\
    \cmidrule(lr){2-3}\cmidrule(lr){4-5}\cmidrule(l){6-6}
    Unit & {min} & {max} & {min} & {max} & {sum} \\
     & \multicolumn{2}{c}{[MW]} & \multicolumn{2}{c}{[\si{\mega\watt\hour}]} & [\si{\mega\watt\hour}] \\
    \midrule
        Super-cap & -10.67 & 480.67 & -0.94 & 0.04 & 0.97 \\
        Flywheel & -50.41 & 756.30 & -3.98 & 0.28 & 4.26 \\
        Battery & -257.03 & 907.28 & -43.27 & 5.87 & 49.14 \\
        DR & -460.40 & 872.53 & -240.56 & 1.66 & 242.22 \\
        Thermal &-793.47 & 1120.63 & -550.47 & 0.08 & 550.55 \\
        Intra-day & 0.00 & 2327.08 & {--} & {--} & {--} \\
    \bottomrule
  \end{tabular}

\end{table}

A one area system with parameters corresponding closely to the central European interconnected power system, given in Table~\ref{tab:sim}, was simulated. A loss of a \SI{1.5}{\giga\watt} plant after \SI{100}{\second} disturbs the system. Figure~\ref{fig:CaseStudy} shows activation of the reserves. It can clearly be seen how different reserves are activated according to the response capabilities, but also how they are relieved and recharged. The SoC returns to the starting SoC for all units. Figure~\ref{fig:CaseStudy_reg} shows system frequency response in blue and AGC activation in green. The system frequency is quickly returned to nominal values, and AGC is soon relieved by the energy bought at intra-day markets. Numerical results are indicated in Table~\ref{tab:results}.

Note, that this is just a exemplary case study to illustrate the here proposed frequency control reserve framework. Many simplifications were made. Especially, all storage units are assumed to be ideal, that is loss-less. However, \cite{Borsche2013GM} showed how losses can be explicitly included in such a framework. There is ample room for improvement on the inter-play between the different units, also, the amount of information that needs to be exchanged should be studied in more detail. Specifically, effects of inaccurate or missing information should be investigated. Nevertheless, it is the belief of the authors that the set-up is very robust against inaccuracies introduced by storage losses and incomplete information.

\section{Conclusion}
To motivate the proposed frequency control reserve framework, economic and technical benefits of allowing energy constrained units to provide reserves were discussed in detail. Specifically, decoupling of control power provision and energy production is found to be economically and environmentally beneficial in a wide range of situations, as it leads to more optimal dispatch results due to the vanishing of must-run generation constraints.

Next, the frequency spectrum of the current control signals and system frequency were analysed. Even today, each frequency regulation service is mainly responsible for a certain part of the spectrum, however the crucial difference to our proposed framework is that in current schemes no guarantees are given on the signal being zero-mean over any period of time -- which is critical for any practical integration of energy storage units for the provision of frequency control reserves.

To enable energy constrained units such as storage systems and \ac{DR} to participate in frequency control, a control reserve framework allowing units to bid for a certain part of the spectrum was introduced. This creates a market and thus a price for parts of the spectrum of both disturbances and reserves. This allows to reimburse units according to their actual benefit to the power system, and penalize those creating the disturbance.

One possible way of splitting the spectrum into arbitrary bands is described, and a case study employing this approach shows how many different technologies are now able to participate in the frequency reserve market. The aggregated response is as fast and smooth as the response of the current frequency control framework, while overall costs might be reduced by allowing more players to participate in the markets.

While this paper motivates the general framework and shows its numerous advantages, there are many details that should be subjected to more scrutiny. This includes the optimal interplay between units providing reserves in different parts of the spectrum, and the required control power for different frequency bands with respect to both every-day operation and contingencies.

\bibliographystyle{IEEEtran}
\bibliography{library}

\end{document}